\documentclass[12pt]{article}
\usepackage{a4wide,feynarts,amsmath,amssymb,graphicx,psfrag}

\newcommand{\ehsp}{\hspace{0.25cm}}

\newcommand{\polv}[3]{\varepsilon_{#1}^{#2}(#3)}

\newcommand{\gev}{\text{GeV}}

\newcommand{\pb}{\text{pb}}
\newcommand{\MM}{{\cal M}}
\newcommand{\Sq}{   {\tilde{q}} }
\newcommand{\Stop}{ {\tilde{t}} }
\newcommand{\Sbot}{ {\tilde{b}} }

\newcommand{\MSUSY}{ M_{\text{SUSY}} }

\newcommand{\sqrts}{\sqrt{\hat{s}}} 

\newcommand{\tb}{\tan\beta}

\begin{document}
\setlength{\unitlength}{1mm}


\begin{titlepage}
\begin{flushright}
{\bf hep-ph/0305321\\
MPI-PhT/2003-25}
\end{flushright}
\vspace{2cm}

\begin{center}
{\Large 
MSSM Higgs bosons associated with high-$p_T$ jets\\[.4cm]
at hadron colliders\\
}
\vspace{2.5cm}
{\large \bf Oliver~Brein \footnote{E-mail: obr@mppmu.mpg.de}} 
{\large and} {\large \bf 
Wolfgang~Hollik \footnote{E-mail: hollik@mppmu.mpg.de}
}  \\[0.8cm]
{\normalsize\em Max-Planck-Institut f\"ur Physik,\\
	F\"ohringer Ring 6, D-80805 M\"unchen, Germany}\\[1cm]
\end{center}
\vspace{2cm}
\begin{abstract}
The cross section for the production of the lightest 
neutral Higgs boson in association
with a high-$p_T$ hadronic jet, calculated in the framework of 
the minimal supersymmetric standard model (MSSM), is presented. 
The expectations for the hadronic cross section 
at the Large Hadron Collider
are discussed using reasonable kinematical cuts. 
In particular the contributions from superpartner loops to
the cross section and their dependence on the parameters of the 
MSSM are investigated and found to be significant.
Comparisons 
show that the 
production rate for $h^0$ + jet in the MSSM can differ widely from  
the corresponding 
standard-model prediction.
\end{abstract}

\vspace*{2cm}

\end{titlepage}


\section{Introduction}

The search for the Higgs boson is a central task at hadron
colliders like the Tevatron and the LHC. 
Most promising Higgs production processes are those with rates high
enough for detection and with a clean signal
that can be separated effectively from the background processes 
\cite{leshouchesHWG,tevatronHWG}.
Especially, the detection of a Standard Model Higgs boson with a mass 
roughly between 100 and 140$\,\gev$ at the LHC 
is rather difficult, because the predominant decays of the Higgs bosons
into $b\bar b$-pairs are swamped by the large QCD two-jet 
background \cite{ATLASTDR}.
Therefore, only in combination with the rare decay of 
the Higgs boson into two photons,
is the inclusive single Higgs boson production 
considered the best search channel 
in this mass range at LHC to date.
The theoretical prediction 
for the inclusive single Higgs production cross section 
has been studied in great detail for the standard model
even including NNLO QCD corrections \cite{inclusive-higgs-sm}.
Alternatively 
and in order to fully explore the Higgs detection capabilities 
of the LHC detectors 
one can investigate more exclusive channels 
like e.g.~Higgs production in association with a high-$p_T$ hadronic jet.
The main advantage of this channel is the richer kinematical structure 
of the events compared to the inclusive single Higgs production.
This allows for refined cuts increasing the signal-to-background ratio.

The process $p p \to H + \text{jet} + X$ in the standard model (SM)
has been studied more than a decade ago by several authors 
\cite{EHSvdB,BaurGlover,CLLR} concentrating on different Higgs decay
signals like  $\tau^+ \tau^-$ in \cite{EHSvdB},
$W^+ W^-$ and $Z Z$ in \cite{BaurGlover} and $b\bar{b}$ in \cite{CLLR}.
Recently, a detailed signal-to-background analysis of 
this Higgs boson production process followed by the Higgs decaying 
into two photons appeared \cite{leshouchesHWG,ADIKSS} with promising
results. 
But, 
in the framework of the minimal supersymmetric standard model (MSSM) 
the theoretical prediction for $p p \to H + \text{jet} + X$ is 
incomplete, missing the contribution of the superpartners, 
like squarks and gluinos,
in contrast to 
the inclusive Higgs production, where the superpartner
contributions are even known at two-loop accuracy in the heavy
squark limit \cite{DDS}.

In this paper we 
present
a complete one-loop result in the 
MSSM for the production of a neutral Higgs boson
in association with a high-$p_T$ hadronic jet  
($p p \to h^0 + \text{jet} + X$).
Furthermore, we explore the
contribution of the previously unknown superpartner loops to the cross
section numerically
for the three benchmark scenarios for the MSSM Higgs search at LEP
\cite{improvedbm}
within regions not
yet excluded by experimental constraints. 
The so-called large-$\mu$ scenario is suitably modified 
to yield Higgs masses not excluded by LEP Higgs searches.
We also compare the MSSM result 
to leading order with the SM result to the same order.
The numerical analyses
are done for LHC energies and parton luminosities.

\section{Partonic Processes}

There are three partonic processes contributing to the hadronic Higgs-plus-jet
signal process: the production of a Higgs boson and a gluon by gluon
fusion or quark-antiquark annihilation ($g g,q \bar{q} \to g h^0$), and of a
Higgs boson accompanied by a quark 
via quark-gluon scattering ($q g \to q h^0$, $\bar q g \to \bar q h^0$).
Our analysis considers 
all three processes, although only gluon fusion and quark-gluon scattering
lead to significant rates at the LHC.

\subsection{Gluon fusion ($g g\to g h^0$)}
\label{gghg-partcs}

In our conventions, 
g$k_{12,3},\,p$ denote the momenta
$\sigma_{1,2,3}$ the helicities, and $a,b,c$ the colour indices
of the particles in the process
\[
 g(k_1,a,\sigma_1) + g(k_2,b,\sigma_2)  
  \rightarrow g(k_3,c,\sigma_3) + h^0(p) \, .
\]
We make use of the partonic kinematical invariants
\begin{align}\label{def-stu}
\hat{s}  = (k_1+k_2)^2\, , \quad 
\hat{t}  = (k_1-k_3)^2\, , \quad
\hat{u}  = (k_1-p)^2 \, ,
\end{align}
obeying the relation
\begin{align}\label{stu-rel}
\hat{s} + \hat{t} + \hat{u} = m_{h^0}^2
\ehsp.
\end{align}
After averaging and summing over the spin and colour degrees of freedom
of the incoming and outgoing particles, respectively, the differential
cross section reads
\begin{align}
\frac{d \sigma_{gg\to g h^0}}{ d \hat{t}} & = \frac{1}{16\pi \hat{s}^2} \:
\frac{1}{4} \sum_{\sigma_1,\sigma_2,\sigma_3 = \pm 1}^{}
\frac{1}{64}
\sum_{a, b, c = 1}^{8}
\big| \MM_{\sigma_1\sigma_2\sigma_3}^{a b c} \big|^2 \, ,
\end{align}
containing the helicity amplitudes
\begin{align}
\MM_{\sigma_1\sigma_2\sigma_3}^{a b c} & =
\polv{\sigma_1}{\mu}{k_1} \polv{\sigma_2}{\nu}{k_2} 
\polv{\sigma_3}{\ast \rho}{k_3} \,
\widetilde{\MM}_{\mu\nu\rho}^{a b c} 
\ehsp, 
\end{align}
where $\polv{\sigma_1}{\mu}{k_1}$, 
      $\polv{\sigma_2}{\nu}{k_2}$ and  
      $\polv{\sigma_3}{\ast \rho}{k_3}$ 
are the polarisation vectors of the gluons.
As a general feature of the amplitude, 
the transversality of gluons gives useful identities 
\begin{align}
k_1^\mu \polv{\sigma_2}{\nu}{k_2} 
\polv{\sigma_3}{\ast \rho}{k_3} \,
\widetilde{\MM}_{\mu\nu\rho}^{a b c}
=
\polv{\sigma_1}{\mu}{k_1} k_2^\nu 
\polv{\sigma_3}{\ast \rho}{k_3} \,
\widetilde{\MM}_{\mu\nu\rho}^{a b c}
= 
\polv{\sigma_1}{\mu}{k_1}
\polv{\sigma_2}{\nu}{k_2}
k_3^\rho \,
\widetilde{\MM}_{\mu\nu\rho}^{a b c}
=
0
\ehsp,  \label{inv}
\end{align}
which allow cross-checks of our one-loop calculation of 
$\widetilde{\MM}_{\mu\nu\rho}^{a b c}$. 

\begin{figure}[tb]
\begin{center}
\input{gghg-qtri}
\input{gghg-qbox}
\caption{\label{gghg-qdiag}
Quark loop graphs for the process $g g \to g h^0$ in leading order.
Feynman graphs with opposite direction of charge flow are not depicted.}
\end{center}
\end{figure}

\begin{figure}[hbt] 
\begin{center}
\input{gghg-sq}
\caption{\label{gghg-sqdiag}
Scalar quark loop graphs for the process $g g \to g h^0$ in leading order.
Feynman graphs with opposite direction of charge flow are not depicted.}
\end{center}
\end{figure}

The process $g g \to g h^0$ is a pure quantum effect, induced at the 
one-loop level. 
The one-loop Feynman graphs contributing to the process divide 
in those which contain a closed quark loop (Figure \ref{gghg-qdiag})
and those which contain a closed loop of virtual squarks 
(Figure \ref{gghg-sqdiag}). 
The quark-loop graphs contain either 
the loop-induced Higgs-gluon-gluon coupling 
with one of the gluons exchanged internally, 
or a box-type topology.
For the squark loop amplitudes the same subdivision applies,
if the fact is taken into account that certain 
$(n-1)$-point loop graphs are connected to certain $n$-point loop graphs
due to the scalar nature of the squarks.
The connection is that certain triangle loop graphs can be obtained
from certain box loop graphs by shrinking one squark line
to a point.
The Feynman graphs for the corresponding Standard Model 
process are analogous to the set of quark loop graphs.

\subsection{Quark-gluon scattering 
	($q g \to q h^0$, $\bar q g \to \bar q h^0$)}
\label{qghq-partcs}
With the same conventions for momenta and helicities as before, 
we consider the processes
\begin{align*}
 g(k_1,a,\sigma_1) + q(k_2,i,\sigma_2)
  & \rightarrow q(k_3,j,\sigma_3) + h^0(p) \;,\\
g(k_1,a,\sigma_1) + \bar q(k_2,i,\sigma_2)
  & \rightarrow \bar q(k_3,j,\sigma_3) + h^0(p) \;.
\end{align*}
The labels $a$ and $\sigma_1 (= \pm 1)$ denote the 
degrees of freedom of colour and helicity of the gluon. 
$a,i,j$ are the colour indices of the gluon and the (anti-)quark. 
The Mandelstam variables are defined 
as in eq.~(\ref{def-stu}) and fulfil the relation (\ref{stu-rel}) 
in the limit of vanishing quark masses. 

\begin{figure}[tb]
\begin{center}
\input{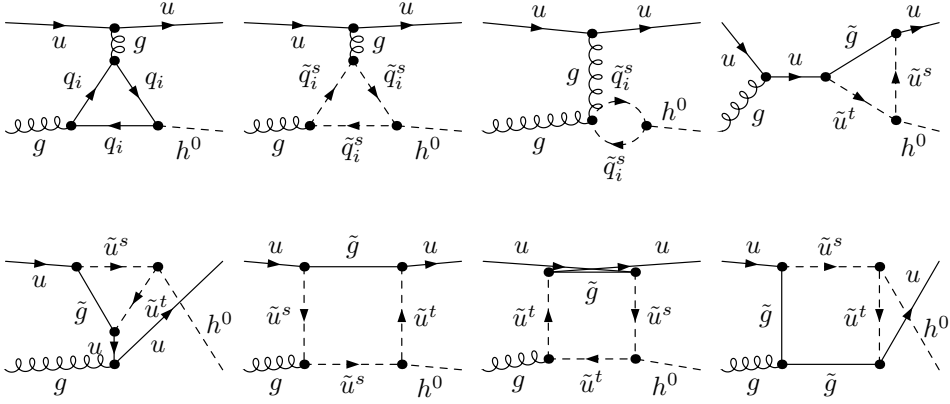}
\caption{\label{ughu-all}
Loop graphs for the process $u g \to u h^0$ in
leading order. 
Feynman graphs with opposite direction of charge flow are not depicted.
For the scattering of the other quarks ($d,c,s$) the graphs look similar.
}
\end{center}
\end{figure}

The differential cross section for quark scattering
averaged over initial 
and summed over final state colour and helicity d.o.f. is given by
\begin{align}
\frac{d \sigma_{qg\to q h^0}}{ d \hat{t}} & = \frac{1}{16\pi \hat{s}^2} \:
\frac{1}{2} 
\sum_{\sigma_1= \pm 1}^{} 
\frac{1}{2}
\sum_{\sigma_2,\sigma_3 = \pm \frac{1}{2}}^{}
\frac{1}{8} 
\sum_{a = 1}^{8} 
\frac{1}{3}
\sum_{i,j = 1}^{3}
\big| \MM_{\sigma_1\sigma_2\sigma_3}^{a i j} \big|^2 \;,
\end{align}
containing the helicity amplitudes
\begin{align}
\MM_{\sigma_1\sigma_2\sigma_3}^{a i j} & =
\polv{\sigma_1}{\mu}{k_1} \,
\widetilde{\MM}_{\mu\: \sigma_2\sigma_3}^{a i j}
\ehsp,
\end{align}
and likewise for anti-quarks.
The polarisation vector of the incoming gluon is denoted by 
$\polv{\sigma_1}{\mu}{k_1}$.
The transversality of the helicity amplitudes (c.f. section \ref{gghg-partcs})
is expressed by the identity,
\begin{align}
k_1^\mu \, \widetilde{\MM}_{\mu\: \sigma_2\sigma_3}^{a i j} & = 0
\end{align}
which has been used as a check of our results .

The Yukawa couplings of the Higgs boson $h^0$ 
to the light quarks ($q = u, d, s, c$) are negligible.
Hence, the leading order amplitude for $q g \to q h^0$ is loop-induced, 
described by the Feynman graphs of ${\cal O} (g_S^3 g_2)$ 
depicted in Figure \ref{ughu-all}.
\footnote{ $g_S$ and $g_2$ are the coupling constants of the strong 
and weak interaction respectively.}
The Feynman graphs divide into those with a closed quark ($q_i$) loop of
triangle-type and those with a loop of virtual 
superpartners (squarks, $\Sq_i^s$, and gluinos, $\tilde g$) 
of triangle- and of box-type.
There are also electroweak contributions $\propto g_S g_2^3$ 
at the one-loop level. We calculated those contributions and found 
their effect on the numerical results for the partonic cross section
to be at most 2~-~3~\% of the QCD contribution. 
They will be 
neglected in the follwwing.

\begin{figure}[tb]
\begin{center}
\input{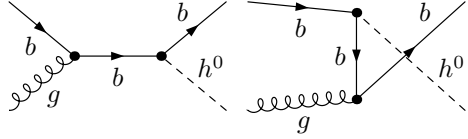}
\caption{\label{bghb-born}
Feynman graphs for the $b$-quark processes in leading order. 
The graphs represent
the amplitude for the process $b g \to b h^0$ if the time axis
points to the right and $b\bar b \to g h^0$  if 
it
points from up to down.
}
\end{center}
\end{figure}

For the $b$-quark, 
the Yukawa-coupling to the Higgs boson cannot be neglected, especially for 
large values of $\tb$, and leads to a tree-level contribution,
as displayed in Figure \ref{bghb-born}. 

The Feynman graphs appearing in the corresponding Standard Model 
process are analogous to the set of quark loop graphs 
in Figure \ref{ughu-all}.
Thus, in the MSSM there appear additional Feynman-graph topologies 
which are not present
in the Standard Model and which may even change the angular 
distribution of the final-state particles compared to the Standard Model
expectation.

\subsection{Quark-antiquark annihilation ($q \bar q \to g h^0$)}
\label{qghg-partcs}

With the same notations for momenta, helicities and colour as above, 
the parton reaction
\begin{align}
q(k_1,i,\sigma_1) +\bar q(k_2,j,\sigma_2)
	& \rightarrow g(k_3,a,\sigma_3) + h^0(p) \;,
\end{align}
has the differential cross section
\begin{align}
\frac{d \sigma_{q\bar q\to g h^0}}{ d \hat{t}} & = \frac{1}{16\pi \hat{s}^2} \:
\frac{1}{4} 
\sum_{\sigma_1,\sigma_2= \pm \frac{1}{2}}^{} 
\frac{1}{9}
\sum_{i,j = 1}^{3}
\sum_{\sigma_3 = \pm 1}^{}
\sum_{a = 1}^{8} 
\big| \MM_{\sigma_1\sigma_2\sigma_3}^{a i j} \big|^2 \;,
\end{align}
containing the helicity amplitudes
\begin{align}
\MM_{\sigma_1\sigma_2\sigma_3}^{a i j} & =
\polv{\sigma_3}{\star \mu}{k_3} \,
\widetilde{\MM}_{\mu\: \sigma_1\sigma_2}^{a i j}
\end{align}
with the polarisation vector $\polv{\sigma_3}{\star \mu}{k_3}$
of the outgoing gluon. 
Analogously to the transversality check of the helicity amplitudes
in the previous two sections, we verified that the helicity
amplitudes vanish if the polarisation vector of the gluon is replaced
by its momentum.

\begin{figure}[tb]
\begin{center}
\input{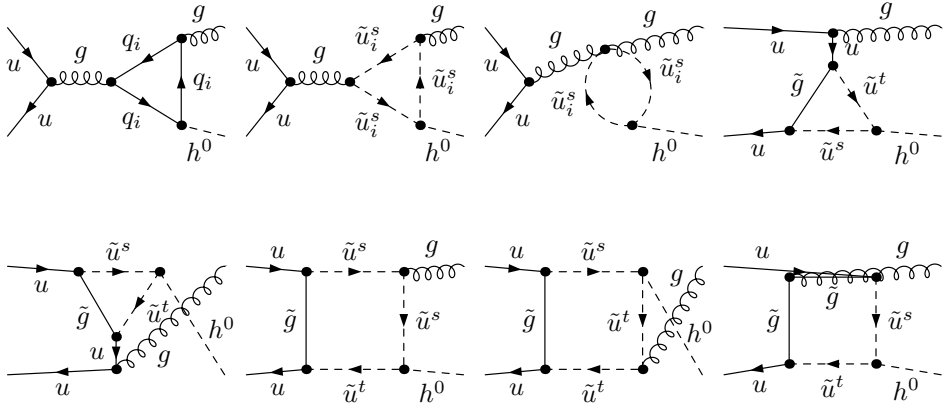}
\caption{\label{uuhg-all}
Feynman graphs for the process $u \bar u \to g h^0$ in
leading order. 
Feynman graphs with opposite direction of charge flow are not depicted.
For the scattering of the other quarks ($d,c,s$) the Feynman graphs 
look similar.
}
\end{center}
\end{figure}

Again, since the Yukawa-couplings of the light quarks to the Higgs boson 
can be neglected, 
the amplitudes for $q\bar q\to g h^0$ ($q = u,d,s,c$) are loop-induced.
The Feynman graphs contributing to these amplitudes 
divide into the ones of Standard Model type,
which contain a quark loop,  
and the others with virtual superpartners, 
squarks and gluinos (see Figure \ref{uuhg-all}).
As in $b$-quark--gluon scattering, the $b$ Yukawa coupling to the Higgs boson
cannot be neglected,
yielding tree-level graphs
obtained by crossing those of Figure \ref{bghb-born}.

\bigskip
\noindent The calculations of the partonic cross sections 
have been performed with the help of 
the computer programs Feyn\-Arts and Form\-Calc \cite{FAFC}.

\section{Hadronic Cross Section} 

The hadronic cross section for $h^0$ + jet production 
with a total hadronic CMS energy 
$\sqrt{S}$ can be written as a convolution \cite{pQCD-lect}
\begin{multline}\label{hadronx}
\sigma  ( A B  \to h^0 + \text{jet} + X) =  
\int_{\tau_0}^{1} d\tau 
\bigg(
\frac{ d{\cal L}^{A B}_{g g} }{ d\tau }
\;\sigma_{g g \to g h^0}(\tau S,\alpha_S(\mu_R)) \\
  + \sum_{q= u,\bar u,\ldots,b,\bar b} \frac{ d{\cal L}^{A B}_{q g} }{ d\tau }
\;\sigma_{q g \to q h^0}(\tau S,\alpha_S(\mu_R))
 +\sum_{q= u,\ldots,b} \frac{ d{\cal L}^{A B}_{q \bar q} }{ d\tau }
\;\sigma_{q \bar q \to g h^0}(\tau S,\alpha_S(\mu_R))
\bigg)
\end{multline}
with the parton luminosity
\begin{align}
\frac{ d{\cal L}^{AB}_{nm} }{ d\tau } & =
        \int_{\tau}^{1} \frac{dx}{x}
        \frac{1}{1+\delta_{nm}} 
        \Big[
        f_{n/A} (x,\mu_F) f_{m/B} (\frac{\tau}{x},\mu_F)
        + f_{m/A} (x,\mu_F) f_{n/B} (\frac{\tau}{x},\mu_F)
        \Big]
        \, ,
\end{align}
where $f_{n/A} (x,\mu_F)$ denotes the density of partons of type $n$
in the hadron $A$ carrying a fraction $x$ of the hadron momentum
at the scale $\mu_F$ and $(A,B)=(p,p)$ for the LHC [and $(p,\bar{p})$
for the Tevatron].
The lower bound of the $\tau$-integration ($\tau_0$) determines
the minimal invariant mass of the parton system, $\hat s_0 = \tau_0 S$.
In our case, $\tau_0$ depends on the kinematical cuts applied 
in order to have high-$p_T$ jets not too close to the beam-pipe.

The angular dependences of the differential cross sections
in the CMS 
of the processes $gg\to g h^0$ and $q g \to q h^0$ (with $m_q =0$)
diverge in the collinear limit, 
i.e. for scattering angle $\hat{\theta} = 0$, or $\hat t = 0$. 
In the gluon fusion process the final state gluon can be collinear to
either of the incoming gluons. 
Therefore, the gluon fusion cross section also diverges 
for $\hat{\theta} = \pi$, or $\hat u = 0$.
These regions of phase space, however,
are excluded when we require a high-$p_T$ jet
originating from the final-state parton.
Specifically, if we impose the condition $k_{3,T} > p_{T,\text{min}}$
for the transverse momentum of the outgoing parton $k_{3,T}$,
one obtains an energy and an angular cut, 
\begin{align}\label{s-cut}
\sqrts & > p_{T,\text{min}} + \sqrt{m_h^2+p_{T,\text{min}}^2} 
	\equiv \sqrt{\tau_0 S} 
\; ,\\
| \cos\hat{\theta} \, | & < 
	\sqrt{1-\frac{4\, \hat{s}\, p_{T,\text{min}}^2}{(\hat{s}-m_h^2)^2}}
	\; .
\end{align}
On top, in order to avoid 
high-$p_T$ partons with very small angles to the 
beam axis in the laboratory frame, $\hat\theta_{\text{lab}}$,
we impose a cut on the pseudo-rapidity of the outgoing parton
$\eta_{3,\text{lab}}$  ($\equiv - \ln \tan (\hat \theta_{\text{lab}}/2)$)
in the laboratory frame, $| \eta_{3,\text{lab}} | < \eta_{\text{max}}$, 
thus constraining the scattering angle in the laboratory 
according to
\begin{align}\label{ang-cut2}
2 \, \arctan e^{-\eta_{\text{max}}} & < \hat\theta_{\text{lab}} 
	< 2 \, \arctan e^{\eta_{\text{max}}}
\; .
\end{align}
This requirement leads to bounds on $\hat{\theta}$, or $\hat t$,
respectively in the CMS frame.
All integrated partonic cross sections
\begin{align}
\sigma_{n m \to m' h^0}(\hat{s},\alpha_S(\mu_R)) & =
\int_{ \hat{t}_{\text{min}} }^{ \hat{t}_{\text{max}} } 
\!\!\! d\hat{t}
\: \frac{d \sigma_{n m \to m' h^0}}{ d \hat{t}} 
	\hspace*{1cm} ({n,m,m'}\in \{q,g\})
\end{align}
are then evaluated by numerical integration over the $\hat{t}$-range
respecting the cuts $k_{3,T} > p_{T,\text{min}}$ 
and $| \eta_{3,\text{lab}} | < \eta_{\text{max}}$.

The numerical evaluation has been carried out with the MRST
gluon distribution functions \cite{MRST} and with the renormalisation
and factorisation scale $\mu_R$ and $\mu_F$ chosen 
both equal to $\sqrts$. 

\section{Numerical Results}

In the following discussion we want to illustrate the
the MSSM predictions for the hadronic process 
$pp \to h^0 + \text{jet} + X$ 
and outline differences between the cases of MSSM Higgs boson $h^0$
and of a standard Higgs particle with the same mass.
In particular we discuss the influence of the diagrams with 
the superpartners squarks and gluinos.
The cuts chosen for the numerical evaluation are chosen as
$p_{T,\text{min}} = 30\,\gev$ 
and $\eta_{\text{max}} = 4.5$, 
which have been used in previous studies in the 
Standard Model~\cite{ADIKSS}.

\subsection{Parameters}\label{parameters}

We
adopt for our discussion the three MSSM 
benchmark scenarios
for the Higgs search at LEP, which
were originally proposed in \cite{improvedbm}.
One of them, the
large-$\mu$ scenario, had to be modified in order to obey
the exclusion limit for the Higgs mass set by the latest
LEP data \cite{LEPHWG}. 
In our discussion we 
vary the common sfermion mass scale $\MSUSY$. 
Therefore, brief specifications of the three scenarios with variable 
sfermion mass scale are given here.
\begin{description}
\item{{\em no-mixing scenario }:} 
The off-diagonal term $X_t$ ($= A_t-\mu\cot\beta$) 
in the top-squark mass matrix is zero,
corresponding to a local minimum of $m_h$ as a function of $X_t$.
The supersymmetric Higgsino mass parameter $\mu$ is set to $- 200 \,\gev$,
the gaugino mass parameter to $M_2 = 200 \,\gev$,
and the gluino mass to
$M_{\tilde g} = 800 \,\gev$ .
When $\tb$ is changed, $A_t$ is changed accordingly to insure $X_t = 0$. 
The settings of the other soft-breaking scalar-quark Higgs couplings 
are $A_b = A_t$ and $A_q = 0$ ($q=u,d,c,s$).

\item{{\em maximal-$m_h$ scenario }:}
$X_t$ is set to $2 \MSUSY$
which yields the maximal value of $m_h$ with respect to stop mixing
effects \footnote{The choice of $X_t$ applies to the Feynman 
diagrammatic calculation of $m_h$ \cite{fd-higgs}.}. 
The other parameters are chosen as in the previous scenario.

\item{{\em large-$\mu$ scenario }:}
The key feature of this scenario is a relatively large value of 
$\mu$ ($= 1000 \,\gev$) which leads to strong radiative corrections 
of the $b$-quark Yukawa coupling to the Higgs boson. 
In modification of the original scenario (where $X_t = -300\,\gev$)
we set $X_t = -900\,\gev$ which
gives a maximal Higgs mass of approximately $124\,\gev$ compared 
to $107\,\gev$ of the original scenario.
The other parameters are: $A_b = 0$, $M_2 = 200 \,\gev$
and $M_{\tilde g} = 400 \,\gev$.
\end{description}

In order to demonstrate
squark effects, we take a common 
sfermion mass scale $\MSUSY$ in all the figures, 
with a moderately 
high value of $400\,\gev$. 
The mass of the CP-odd Higgs $m_A$, the ratio of Higgs vacuum
expectation values $\tan\beta = v_2/v_1$,
and the common sfermion mass scale $\MSUSY$
are varied within bounds 
from direct Higgs and squark searches (c.f. \cite{pdg2002}). 
Additionally, in all scenarios, mixing in
the squark sector respects the bounds on additional non-SM 
contributions to the electroweak $\rho$-parameter \cite{rho-param}.
For the electroweak parameters we use the values 
given in \cite{pdg2002}, and
for the strong coupling constant $\alpha_S(\mu_R)$, we use the formula 
including the two-loop QCD corrections for $n_f=5$ with 
$\Lambda^5_{QCD}=170$ MeV which can be found also in~\cite{pdg2002}.
%
Effects of non-zero widths of the quarks
and squarks in the loops are not taken into account
because their widths are considerably smaller
than their masses and are expected to change the results only little
\footnote{A study of non-zero width effects for a 
similar process is documented in \cite{gghw}.}.

\begin{table}[tb]
\begin{center}
\begin{tabular}[7]{l|r|r|r|r|r|r|}
Scenario & \multicolumn{2}{|c|}{no mixing}
	 & \multicolumn{2}{|c|}{$m_h^{\text{max}}$}
	 & \multicolumn{2}{|c|}{large $\mu$}\\
 	 & \multicolumn{2}{|c|}{\small ($m_A = 400\,\gev$)}
         & \multicolumn{2}{|c|}{\small ($m_A = 400\,\gev$)}
         & \multicolumn{2}{|c|}{\small ($m_A = 400\,\gev$)}\\
\hline
$\tb$ 	&\multicolumn{1}{|c|}{6}&\multicolumn{1}{|c|}{30}
	& \multicolumn{1}{|c|}{6}&\multicolumn{1}{|c|}{30}
	& \multicolumn{1}{|c|}{6}&\multicolumn{1}{|c|}{30}\\
\hline
$\MSUSY$ [GeV] & {\small 400 } & {\small 400} & \small 400 
	& \small 400 & \small 400 & \small 400 \\
\hline
$m_h$ [GeV] & \small 100.2 & \small 104.5 & \small 120.2 & \small 123.9 
	& \small 121.0 & \small 123.3\\
\hline
\hline
$\sigma^{\text{MSSM}}_{\text{all}}$ 
	& 7.50 & 6.96 & 3.89 & 3.65 & 2.12 & 1.51\\
 (  $\sigma^{\text{MSSM}}_{\text{all, no SP}}$ )
	& (6.49) & (6.10) & (5.04) & (4.84) & (4.89)  & (4.78) \\
\hline
\hline
$\sigma^{\text{MSSM}}_{(gg\to h g)}$ 
	& 4.81 & 4.44 & 2.46 & 2.31 & 1.38 & 0.97\\
\hline
$\sigma^{\text{MSSM}}_{(q g\to h q, q=u,\bar u,d,\bar d,c,\bar c,s,\bar s)}$ 
	& 2.32 & 2.19 & 1.17 & 1.11 & 0.65 & 0.46 \\
\hline
$\sigma^{\text{MSSM}}_{(q g\to h q, q=b,\bar b)}$ 
	& 0.29 & 0.26 & 0.20 & 0.18 & 0.07 & 0.06\\
\hline
$\sigma^{\text{MSSM}}_{(b \bar b\to h g)}$ 
	& 0.08 & 0.07 & 0.06 & 0.05 & 0.02 & 0.02 \\
\hline
$\sigma^{\text{MSSM}}_{(q \bar q\to h g, q=u,d,c,s)}$ 
	& 0.03 & 0.03 & 0.02 & 0.02 & 0.01 & 0.01 \\
\hline
\hline
$\sigma^{\text{SM}}_{\text{all}}$ 
	& 6.27 & 5.92 & 5.07 & 4.85 & 4.88 & 4.75\\
\hline
\hline
$\sigma^{\text{SM}}_{(gg\to H g)}$ 
	& 4.07 & 3.83 & 3.35 & 3.20 & 3.32 & 3.22\\
\hline
$\sigma^{\text{SM}}_{(q g\to H q, q=u,\bar u,d,\bar d,c,\bar c,s,\bar s)}$ 
	& 1.86 & 1.77 & 1.49 & 1.43 & 1.46 & 1.43\\
\hline
$\sigma^{\text{SM}}_{(q g\to H q, q=b,\bar b)}$ 
	& 0.23 & 0.21 & 0.15 & 0.14 & 0.05 & 0.05\\
\hline
$\sigma^{\text{SM}}_{(b \bar b \to H g)}$
        & 0.08 & 0.08 & 0.06 & 0.05 & 0.02 & 0.02\\
\hline
$\sigma^{\text{SM}}_{(q \bar q \to H g, q=u,d,c,s)}$
	& 0.03 & 0.03  & 0.03 & 0.03 & 0.03 & 0.03\\
\hline
\end{tabular}
\end{center}
\caption{\label{table-cs} MSSM and SM predictions 
for the hadronic cross section, divided in the contributions from
different partonic processes.
}
\end{table}

\subsection{Discussion}

The Higgs-boson couplings to quarks and squarks consist entirely
of terms weighted by either $\sin\alpha$ or $\cos\alpha$
(see Appendix \ref{couplings}), with
$\alpha$ being the mixing angle between the neutral CP-even Higgs fields.
Thus, the cross section's dependence on $m_A$ and $\tan\beta$
can be easily inferred from the dependence 
of $\alpha$ on those parameters and the simple 
$\tan\beta$-dependence
of the Higgs couplings to quarks and squarks
(see eqs. (\ref{qqh}) -- (\ref{sb2sb2h})). 
To evaluate the Higgs mass $m_h$ and mixing angle $\alpha$
we use the results of the two-loop Feynman 
diagrammatic calculation 
in the effective-mixing-angle approximation \cite{fd-higgs}, 
where the effective mixing angle $\alpha$ is given by
\begin{align}\label{alpha}
\tan\alpha & = \frac{-(m_A^2 + m_Z^2)\sin\beta \cos\beta 
	- \hat\Sigma_{\phi_1 \phi_2}(0)}
	{m_Z^2 \cos^2\beta + m_A^2 \sin^2\beta
	-\hat\Sigma_{\phi_1}(0) -m_h^2}
\;.
\end{align} 

Figures \ref{ma-hjet}(a)--(f) and \ref{tb-hjet}(a)--(f)
show the results for the hadronic cross section as functions
of $m_A$ and $\tb$ for exemplary values of $\tb$ (6 and 30) 
and $m_A$ (100 and 400 $\gev$) respectively. In all 
plots several contributions to the cross section 
are displayed separately.

Especially  for large $\tb$, the $b$-quark Yukawa coupling is enhanced,
and the partonic processes $b(\bar b) g \to h^0 b(\bar b)$
and $b\bar b \to h^0 g$ are dominated by the tree-level amplitude
of Figure~\ref{bghb-born}.
These tree-level processes are included via the $b$-quark distribution
in the proton, using
a running
$b$-quark mass in the Yukawa coupling, which is known to
take into account the most important effect 
of the NLO corrections \cite{DSSW}.
The thick solid lines in Figures \ref{ma-hjet}(a)--(f) and
\ref{tb-hjet}(a)--(f) show the 
complete cross section in the MSSM, based on the
partonic processes gluon fusion, quark--gluon scattering,
and quark--anti-quark annihilation including the $b$-quark tree-processes.
In order to visualize the size of the $b$-quark effects,
we neglect the $b$-quark distribution
in the proton and use only the effectively loop-induced processes 
for the light quarks ($u,\bar u, d, \bar d,c,\bar c, s, \bar s$),
as illustrated in Figure \ref{ughu-all}.
The cross section for this case is displayed by
the thick dashed lines in Figures \ref{ma-hjet}(a)--(f) and
\ref{tb-hjet}(a)--(f).

In order to demonstrate the influence of the virtual superpartners 
on the result
the thin solid and dashed lines show the hadronic cross section 
obtained by neglecting 
all diagrams with virtual superpartners 
in the calculation
of the corresponding partonic processes.

Next we discuss the behaviour of the cross section in the three 
benchmark scenarios.

\subsubsection*{No-mixing scenario}

The behaviour of the hadronic cross section in the no-mixing
scenario can be inferred from
Figures \ref{ma-hjet}(a) and \ref{ma-hjet}(b), 
which show the $m_A$-dependence
for $\tb = 6$ and $\tb =30$ 
and from Figures \ref{tb-hjet}(a) and \ref{tb-hjet}(b),
which show the $\tb$-dependence for $m_A = 100\,\gev$ and $m_A = 400\,\gev$.

In this scenario the masses of the squarks are essentially
independent of the parameters of the MSSM Higgs sector.
This is because the $\tb$-dependent mixing of squarks
is either zero by definition (in the scalar top sector) 
or very small (for all other squarks) and $m_A$ does not enter
the squark sector at all in leading order.
Therefore, the variation of the cross section with $m_A$ and $\tb$
is determined by the variation of the mass and the couplings of $h^0$.
For small $m_A$ ($< 150\,\gev$) and especially for large $\tb$, 
the $b$-quark Yukawa coupling 
is strongly enhanced compared to the top-Higgs coupling
(see eqs. (\ref{qqh}),(\ref{sb1sb1h}) and (\ref{sb2sb2h})).
Thus, in this parameter range 
the $b$-quark processes dominate the hadronic cross section,
and also the loop-induced processes are dominated by the $b$-quark loops.
At large $m_A$ ($> 200\,\gev$) the coupling of the $b$--Higgs
coupling is much smaller than the top--Higgs coupling and therefore 
the loop-induced processes, mainly the top-quark loops,
dominate the hadronic cross section. In this parameter range 
the superpartners contribute about 15\% of the complete result
(see Table \ref{table-cs}).  
For large $m_A$ the flat behaviour of the cross section in
Figure \ref{tb-hjet}(b) illustrates, that relevant
contributions only come from loops with virtual up-type
quarks and squarks, which couple $\propto 1/\sin\beta$ 
to the $h^0$, while down-type quarks and squarks couple 
$\propto 1/\cos\beta$ (see eqs. (\ref{qqh}) -- (\ref{st2st2h})).

Figure \ref{msusy-plot}(a) shows the dependence of the cross section 
on the sfermion-mass scale $\MSUSY$ for $m_A = 200\,\gev$ 
and $\tb = 6$. The relative difference between 
the full result and the one without superpartner loops stays 
above 10\% for $\MSUSY$ below $500\,\gev$.

\subsubsection*{Maximal-$m_h$ scenario}

The $m_A$-dependence of the hadronic cross section in the 
maximal-$m_h$ scenario is shown in 
Figures \ref{ma-hjet}(c) ($\tb = 6$) and \ref{ma-hjet}(d) ($\tb =30$),
and the $\tb$-dependence in 
Figures \ref{tb-hjet}(c) ($m_A = 100\,\gev$) 
and \ref{tb-hjet}(d) ($m_A = 400\,\gev$).

This scenario is rather similar to the no-mixing scenario
as far as the variation of the cross section with $m_A$ and $\tan\beta$
is concerned. The $m_A$- and $\tb$-dependences arise almost 
entirely through the dependence of $h^0$ mass and couplings, while
the squark masses are essentially constant. 
But there is one crucial difference: the sign of the leading squark-loop 
contributions is opposite to the no-mixing case, which results in
a suppression of the full result instead of an enhancement. 
This is because the $\Stop$ mixing
angle $\theta_{\tilde t}$ is approximately $\pi/4$
in the maximal-$m_h$ scenario.
Therefore, the leading behaviour of the    
squark-loop amplitude is governed by the terms proportional to     
$m_t A_t \cos\alpha / \sin\beta$
in the
coupling of the $h^0$ to the lighter top squark (see eq. (\ref{st1st1h})).
This yields a destructive interference with the quark loops.
For $m_A = 400\,\gev$, and almost independent of $\tb$,
the full result is reduced by about 
24\% compared to the result with quark loops only (see Table \ref{table-cs}).

Figure \ref{msusy-plot}(a) shows the dependence of the cross section 
on the sfermion mass scale $\MSUSY$ for $m_A = 200\,\gev$ 
and $\tb = 6$.

\subsubsection*{Large-$\mu$ scenario}

The $m_A$-dependence of the hadronic cross section in the
large-$\mu$ scenario is shown in 
Figures \ref{ma-hjet}(e) ($\tb = 6$) and \ref{ma-hjet}(f) ($\tb =30$)
and the $\tb$-dependence in 
Figures \ref{tb-hjet}(e) ($m_A = 100\,\gev$)
and \ref{tb-hjet}(f) ($m_A = 400\,\gev$).

The large-$\mu$ scenario
shows the most pronounced superpartner contribution.
The superpartner loops interfere destructively with 
the quark loops, except when $\tan\beta$ is large
and $m_A$ is small simultaneously.
In the large-$\mu$ scenario, mixing in the $\Sbot$ sector
increases with $\tb$, resulting in large mixing 
($\theta_{\Sbot} \approx 45^\circ$) for $\tb \geq 30$.
Therefore, 
for large $\tb$,
the terms proportional to $m_b \mu \sin2\theta_{\Sbot}$ 
in the $\Sbot$-Higgs couplings are 
dominant, followed by
the terms proportional to $m_t A_t \sin2\theta_{\Stop}$ in 
the $\Stop$-Higgs couplings (see eqs. (\ref{st1st1h}) to (\ref{sb2sb2h})).

For large $m_A$ the decrease of the cross section lies between 50\% and 70\% 
(see Table \ref{table-cs}).
Figure \ref{msusy-plot}(b) shows the dependence of the cross section 
on the sfermion-mass scale $\MSUSY$ for $m_A = 200\,\gev$ 
and $\tb = 6$. The large negative interference between quark and superpartner
loops vanishes rapidly with rising $\MSUSY$. Unlike for the other 
two scenarios the superpartner contribution is already negligible
for $\MSUSY > 600\,\gev$.

\subsubsection*{Comparison with the Standard Model}

The thin dot-dashed line 
in Figures \ref{ma-hjet}(a)--(f) and \ref{tb-hjet}(a)--(f) 
indicates the hadronic cross section in the Standard Model 
with the mass of the Standard-Model Higgs boson 
chosen equal to $m_h$ in the MSSM.
In this way, we can discuss the difference 
between the two predictions as a function of the MSSM parameters.
Figures \ref{ma-hjet}(a)--(f) show the decoupling behaviour
in the MSSM Higgs sector with rising $m_A$. 
The MSSM prediction
where virtual superpartners are neglected in the calculation (thin solid 
lines), which corresponds essentially to the case of decoupling 
superpartners, approaches the SM prediction closely with rising $m_A$.
In this regime $h^0$
behaves like the Standard Model Higgs boson. 
However, the full MSSM prediction
(thick solid lines) shows a rather large departure from the Standard Model
prediction for 
lower sfermion masses, as in the MSSM parameter scenarios we chose.

In Figure \ref{sm-mssm-comp} we display the relative difference
between the MSSM and the SM prediction of the hadronic cross section 
plotted versus $m_A$ and $\tan\beta$.
The parameters of the maximal-$m_h$ scenario with 
$\MSUSY = 400\,\gev$ are chosen as an example.
For this moderate value of $\MSUSY$ the MSSM prediction for the 
hadronic cross section is more than 20\% below the SM result in the 
whole area of the $m_A$-$\tan\beta$ plane displayed.

\section{Conclusions}

The production of a neutral Higgs boson accompanied by a 
high-$p_T$ jet is considered advantageous for Higgs boson detection
even though the rate is lower than for totally inclusive single Higgs boson
production.
Refined cuts allow to increase the signal-to-background
ratio compared to the inclusive production.
We calculate the hadronic cross section for Higgs-plus-jet production
with the full set of 
MSSM Feynman graphs at leading-order QCD.
We find a quite substantial cross section ranging from about $0.3\,\pb$ 
to $300\,\pb$ depending on the MSSM parameter scenario. 
Thus, this process
might be detectable at the LHC even if 
the $\gamma\gamma$-decay channel of the 
Higgs is considered (see e.g. \cite{leshouchesHWG,ADIKSS}).
The contribution from superpartner loops to
the cross section and its dependence on the parameters of the MSSM
turns out to be significant.
We provide a FORTRAN code for general use.

\bigskip\bigskip
\noindent {\em Acknowledgement.} 

We thank Georg Weiglein 
for bringing this topic to our attention.
This work was supported in part 
by the European Community's Human Potential Programme under contract
HPRN-CT-2000-00149 ``Physics at Colliders''.


\appendix

\section*{Appendix}

\section{Higgs couplings}\label{couplings}

\begin{align}\label{qqh}
g_s[ h^0 t t ] & =
        -g_2\:\frac{m_t}{2m_W}\:\frac{\cos\alpha}{\sin\beta}
\; ,
&
g_s[ h^0 b b ] & =
        +g_2\:\frac{m_b}{2m_W}\:\frac{\sin\alpha}{\cos\beta}
\; .
\end{align}

\begin{multline}\label{st1st1h}
g[ h^0 \Stop_1 \Stop_1 ] = g_2\,
\Big[  \frac{\cos\alpha}{\sin\beta} 
	\Big( \frac{m_t A_t}{2 m_W} \sin 2\theta_{\Stop}
		- \frac{m_t^2}{m_W}
	\Big) 
	+ \frac{\sin\alpha}{\sin\beta}
	\Big( 
		\frac{m_t \mu}{2 m_W} \sin 2\theta_{\Stop}
	\Big) \\
	- \sin(\alpha+\beta) 
        \Big( \frac{m_Z (5 - 8 c_w^2)}{6 c_w} \cos^2\theta_{\Stop} 
		-\frac{2 m_Z s_w^2}{3 c_w}
        \Big)
\Big]
\; ,
\end{multline}

\begin{multline}\label{st2st2h}
g[ h^0 \Stop_2 \Stop_2 ] = g_2\,
\Big[ \frac{\cos\alpha}{\sin\beta}
        \Big( -\frac{m_t A_t}{2 m_W} \sin 2\theta_{\Stop}
                - \frac{m_t^2}{m_W}
        \Big)
        + \frac{\sin\alpha}{\sin\beta}
        \Big(
               -  \frac{m_t \mu}{2 m_W} \sin 2\theta_{\Stop}
        \Big) \\
        - \sin(\alpha+\beta)
        \Big( - \frac{m_Z (5 - 8 c_w^2)}{6 c_w} \cos^2\theta_{\Stop} 
		+ \frac{m_Z (1 -4 s_w^2) }{6 c_w}
        \Big)  
\Big]
\; ,
\end{multline}

\begin{multline}\label{sb1sb1h}
g[ h^0 \Sbot_1 \Sbot_1 ] = g_2\,
\Big[ \frac{\sin\alpha}{\cos\beta}
	\Big( - \frac{m_b A_b}{2 m_W} \sin 2\theta_{\Sbot}
	 	+ \frac{m_b^2}{m_W} 
	\Big)
	+ \frac{\cos\alpha}{\cos\beta}  
	\Big(- \frac{m_b \mu}{2 m_W} \sin 2\theta_{\Sbot}
	\Big) \\
	- \sin(\alpha+\beta)
	\Big( \frac{m_Z (4 c_w^2 - 1 )}{6 c_w} \cos^2\theta_{\Sbot}
		+ \frac{m_Z s_w^2}{3 c_w}
	\Big)
\Big]
\; ,
\end{multline}

\begin{multline}\label{sb2sb2h}
g[ h^0 \Sbot_2 \Sbot_2 ] = g_2\,
\Big[  \frac{\sin\alpha}{\cos\beta}
	\Big( \frac{m_b A_b}{2 m_W} \sin 2\theta_{\Sbot}
		+ \frac{m_b^2}{m_W} 
	\Big)
	+ \frac{\cos\alpha}{\cos\beta}
	\Big( \frac{m_b \mu}{2 m_W} \sin 2\theta_{\Sbot}
	\Big) \\
	- \sin(\alpha+\beta)
	\Big( - \frac{m_Z (4 c_w^2 - 1 )}{6 c_w} \cos^2\theta_{\Sbot} 
		+ \frac{m_Z (1 + 2 c_w^2)}{6 c_w}
	\Big)
\Big]
\; ,
\end{multline}


\newpage

\begin{figure}[hbt]
\begin{center}
  \psfrag{SIGMAPP}[c][c]{$\sigma( pp \to h^0 + \text{Jet}) \; [\pb]$}
  \psfrag{TANB06}{\scriptsize $ \tan\beta = 6$}
  \psfrag{TANB30}{\scriptsize $ \tan\beta = 30$}
  \psfrag{SM}[r][r]{\tiny SM}
  \psfrag{MSSM-ALL}[r][r]{\tiny MSSM,$\,$total}
  \psfrag{MSSM-QUARK}[r][r]{\tiny MSSM,$\,$no SP}
  \psfrag{MSSM-no b}[r][r]{\tiny MSSM,$\,$no $b$}
  \psfrag{MSSM-no b, quark}[r][r]{\tiny MSSM,$\,$no SP,$\,$no $b$}
  \psfrag{NOMIX}[l][l]{\scriptsize no mixing}
  \psfrag{MHMAX}[l][l]{\scriptsize maximal $m_h$}
  \psfrag{LMU2}[l][l]{\scriptsize large $\mu$}
  \psfrag{MA}[c]{$m_A \; [\gev]$}
  \psfrag{(a)}[c][c]{\bf (a)}
  \psfrag{(b)}[c][c]{\bf (b)}
  \psfrag{(c)}[c][c]{\bf (c)}
  \psfrag{(d)}[c][c]{\bf (d)}
  \psfrag{(e)}[c][c]{\bf (e)}
  \psfrag{(f)}[c][c]{\bf (f)}
  \vspace*{-.3cm}
  \hspace*{-.5cm}
   \resizebox*{1.\width}{1.\height}{
        \includegraphics*{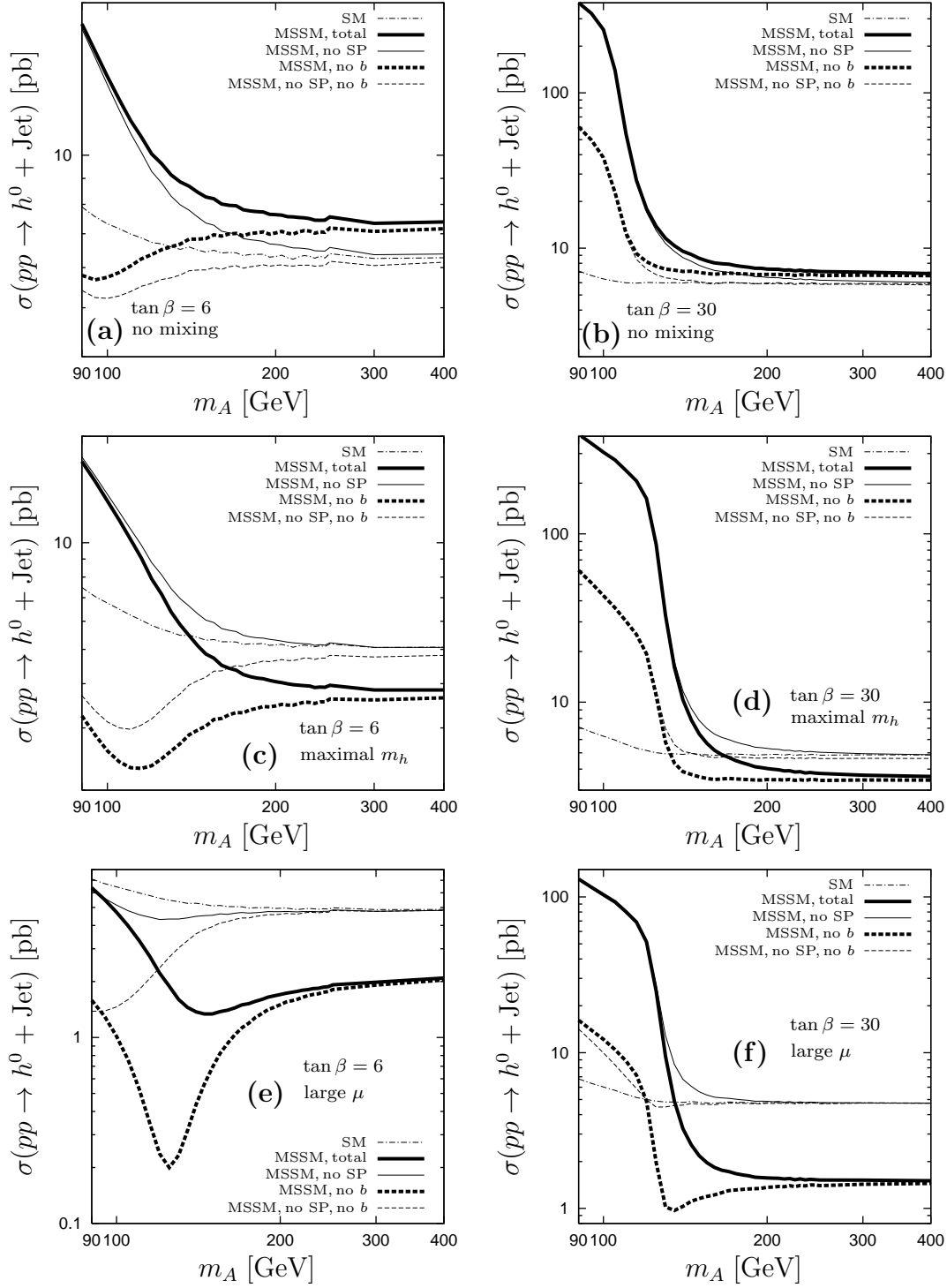}}
    \caption{\label{ma-hjet}
	Hadronic cross section for the process $pp\to h^0 +\text{jet}$
	as a function of $m_A$. The results of the three benchmark
	scenarios with $\MSUSY = 400\,\gev$
	are displayed for $\tb = 6$ and 30. 
	}
  \end{center}
\end{figure}

\begin{figure}[hbt]
\begin{center}
  \psfrag{SIGMAPP}[c][c]{$\sigma( pp \to h^0 + \text{Jet}) \; [\pb]$}
  \psfrag{MA100}{\scriptsize $m_A = 100\,\text{GeV} $}
  \psfrag{MA400}{\scriptsize $m_A = 400\,\text{GeV} $}
  \psfrag{SM}[r][r]{\tiny SM}
  \psfrag{MSSM-ALL}[r][r]{\tiny MSSM,$\,$total}
  \psfrag{MSSM-QUARK}[r][r]{\tiny MSSM,$\,$no SP}
  \psfrag{MSSM-no b}[r][r]{\tiny MSSM,$\,$no $b$}
  \psfrag{MSSM-no b, quark}[r][r]{\tiny MSSM,$\,$no SP,$\,$no $b$}
  \psfrag{NOMIX}[l][l]{\scriptsize no mixing}
  \psfrag{MHMAX}[l][l]{\scriptsize maximal $m_h$}
  \psfrag{LMU2}[l][l]{\scriptsize large $\mu$}
  \psfrag{TANB}[c]{$\tan\beta$}
  \psfrag{(a)}[c][c]{\bf (a)}
  \psfrag{(b)}[c][c]{\bf (b)}
  \psfrag{(c)}[c][c]{\bf (c)}
  \psfrag{(d)}[c][c]{\bf (d)}
  \psfrag{(e)}[c][c]{\bf (e)}
  \psfrag{(f)}[c][c]{\bf (f)}
  \vspace*{-.3cm}
  \hspace*{-.5cm}
   \resizebox*{1.\width}{1.\height}{
        \includegraphics*{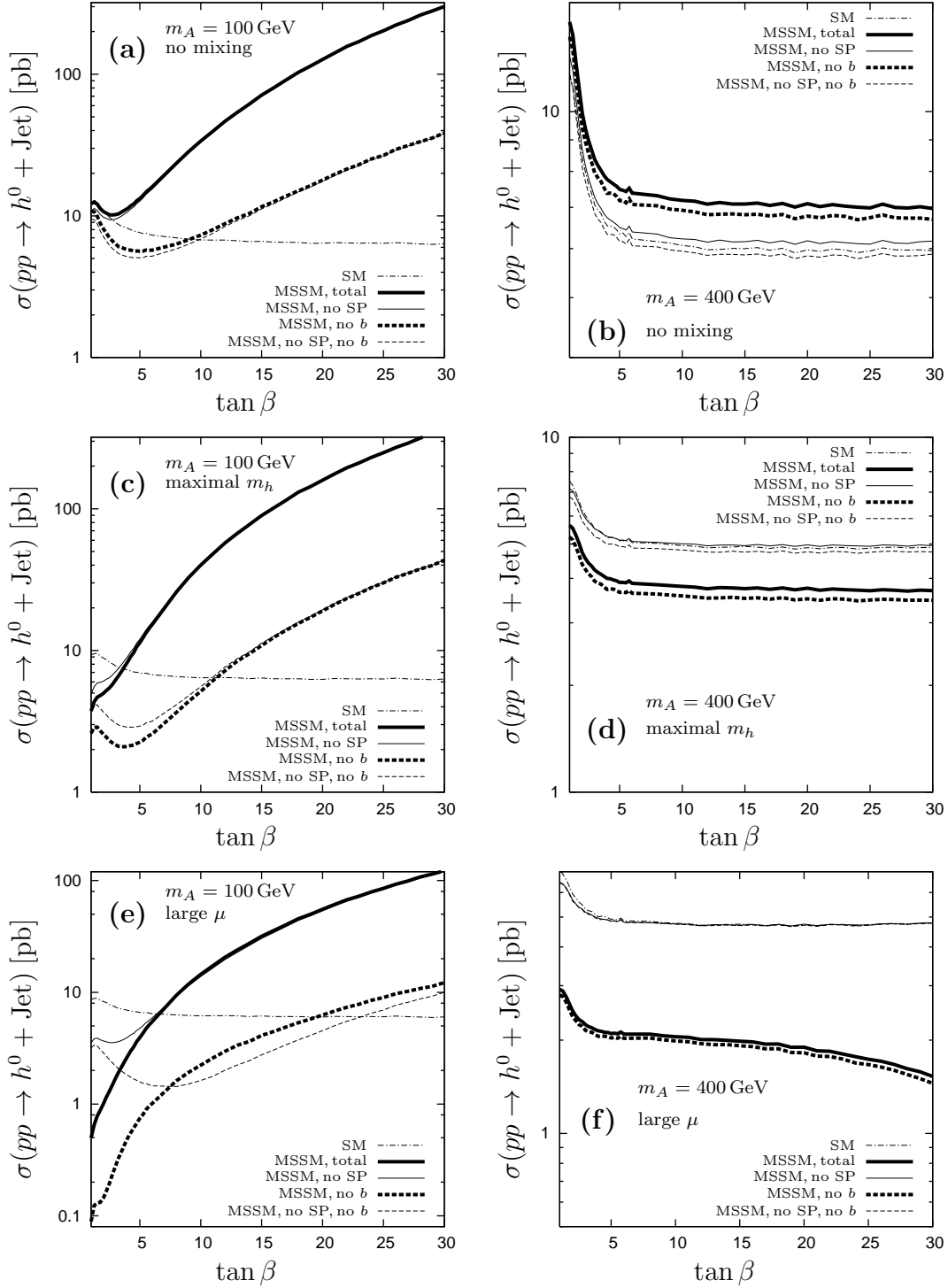}}
    \caption{\label{tb-hjet}
	Hadronic cross section for the process $pp\to h^0 +\text{jet}$
	as a function of $\tb$. The results of the three benchmark
	scenarios with $\MSUSY = 400\,\gev$
	are displayed for $m_A = 100\,\gev$ and $400\,\gev$. 
	}
  \end{center}
\end{figure}

\begin{figure}[bh]
\begin{center}
  \psfrag{SIGMAPP}[c][c]{$\sigma( pp \to h^0 + \text{Jet}) \; [\pb]$}
  \psfrag{MA200}[l][l]{$\scriptstyle m_A = 200\,\text{GeV} $}
  \psfrag{TANB06}[l][l]{$\scriptstyle \tan\beta = 6$}
  \psfrag{MSUSY}[c][b][2]{\tiny $\MSUSY [\gev]$}
  \psfrag{(a)}[c][c]{\bf (a)}
  \psfrag{(b)}[c][c]{\bf (b)}
\psfrag{NOMIX}[r][r]{\tiny no-mixing}
\psfrag{NOMIX-QUARK}[r][r]{\tiny no-mixing, no SP}
\psfrag{MHMAX}[r][r]{\tiny maximal-$m_h$}
\psfrag{MHMAX-QUARK}[r][r]{\tiny maximal-$m_h$, no SP}
\psfrag{LMU2}[r][r]{\tiny large-$\mu$}
\psfrag{LMU2-QUARK}[r][r]{\tiny large-$\mu$, no SP}
\resizebox*{1.1\width}{1.1\height}{\includegraphics*{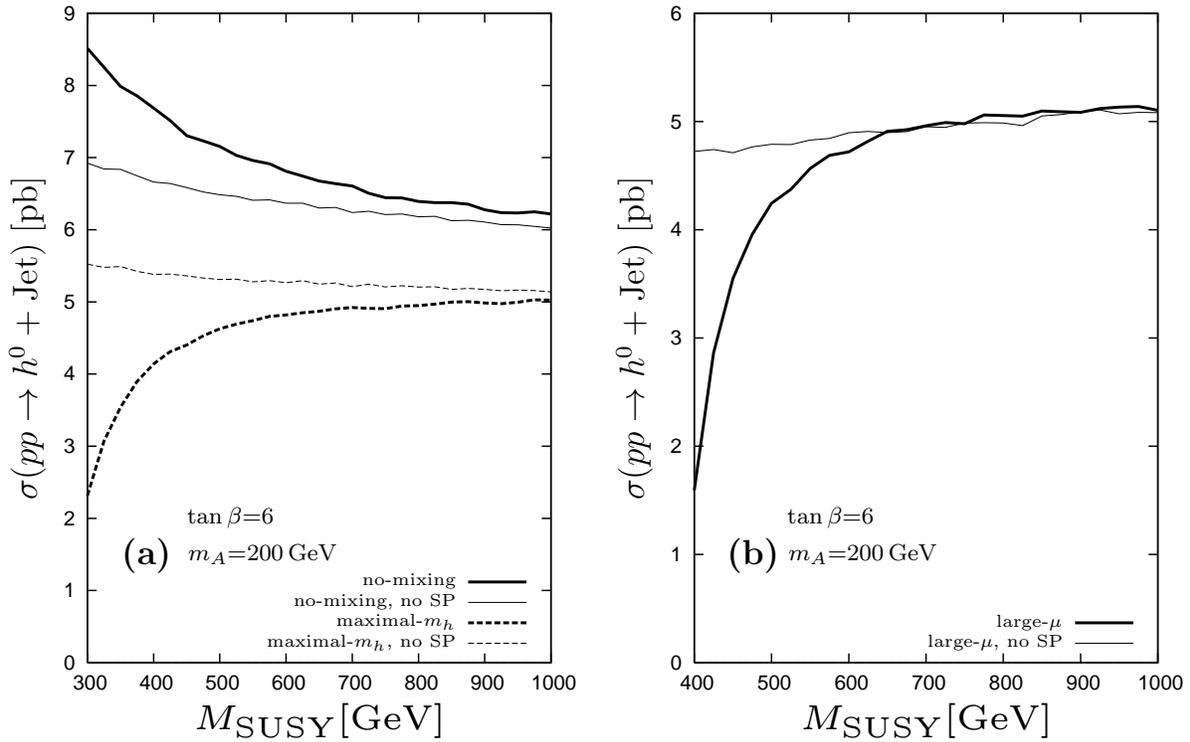}}
    \caption{\label{msusy-plot} 
	Hadronic cross section for
	Higgs plus jet production versus $\MSUSY$.
	$m_A = 200\,\gev$ and $\tan\beta = 6$. The cross section
	prediction is shown for 
  	the no-mixing (solid lines) and 
	maximal-$m_h$ scenario (dashed lines) 
	in panel (a) and 
	for the large-$\mu$ scenario in panel (b).
        Thick lines indicate the full result, while thin lines
	are obtained by leaving out all Feynman 
	graphs with superpartner loops in the calculation.
            }
  \end{center}
\end{figure}

\begin{figure}[bh]
\begin{center}
\psfrag{TANB}{$\tan\beta$}
\psfrag{MA}{$m_A \; [\gev]$}
\psfrag{DELTA}[lb]{$\delta [\%]$}
\psfrag{RANGE1}[l]{$-26\% < \delta < -24\%$}
\resizebox*{1.3\width}{1.3\height}{\includegraphics*{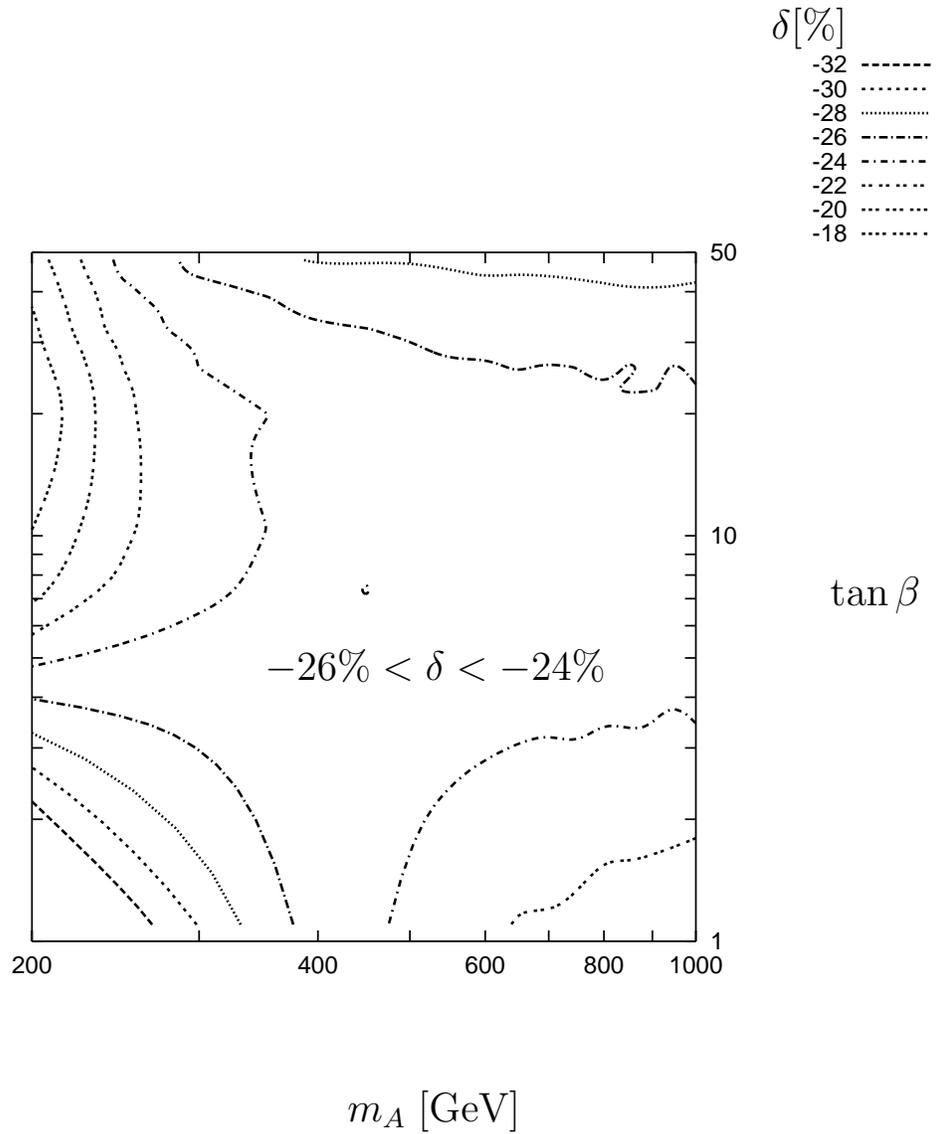}}
    \caption{\label{sm-mssm-comp} 
	Relative difference $\delta$ between the MSSM result and 
	the corresponding SM result (with identical Higgs mass)
	for the hadronic cross section for
        Higgs plus jet production. The dependence of 
	$\delta = (\sigma^{\text{MSSM}}-\sigma^{\text{SM}})/\sigma^{\text{SM}}$ 
	on $m_A$ and $\tan\beta$ is indicated by contour lines. 
        Here the maximal-$m_h$ scenario with $\MSUSY = 400\,\gev$ is chosen.
            }
  \end{center}
\end{figure}


\end{document}